\documentclass{sig-alternate} 

\usepackage{amsmath}
\usepackage{comment}
\usepackage{graphicx}
\usepackage{cite}
\usepackage{url}
\usepackage{epsfig}
\usepackage{varioref}
\newcommand{\choice}[1]{\ensuremath{v^*_{#1}}}

\begin{document}

\title{An Event Based Approach to Situational Representation}

\numberofauthors{6} 

\author{
%% e-mail address with \email.
%\alignauthor \hspace*{-1.2in}\mbox{Naveen Ashish ~~ Dmitri V. Kalashnikov ~~ Sharad Mehrotra ~~ Nalini Venkatasubramanian}\\
 \alignauthor \hspace*{-2.2in}\mbox{Naveen Ashish ~~ Dmitri V. Kalashnikov ~~ Sharad Mehrotra ~~ Nalini Venkatasubramanian}\\
\vspace*{1em}
\mbox{\em Calit2 and ICS Department}\\
\mbox{\em University of California, Irvine}\\
\mbox{\em Irvine, CA 92707, USA}\\
\mbox{ashish@ics.uci.edu}}
\date{}
\maketitle

\begin{abstract} 

Many application domains require representing interrelated real-world activities and/or evolving physical phenomena. In the crisis response domain, for instance, one may be interested in representing the state of the unfolding crisis (e.g., forest fire), the progress of the response activities such as evacuation and traffic control, and the state of the crisis site(s). Such a situation representation can then be used to support a multitude of applications including situation monitoring, analysis, and planning. In this paper, we make a case for an {\em event} based representation of situations where events are defined to be domain-specific significant occurrences in space and time. We argue that events offer a unifying and powerful abstraction to building situational awareness applications. We identify challenges in building an {\em Event Management System (EMS)} for which traditional data and knowledge management systems prove to be limited and suggest possible directions and technologies to address the challenges.

\end{abstract}

\section{Introduction}
\label{sec:intro}

A large number of applications in different domains, including the emergency response and disaster management domain that motivates our work, require capturing and representing information about real-world situations as they unfold. Such situations may correspond to interrelated real-world activities or evolving physical phenomena. Such situational data is usually extracted from multi-sensory data of different (and possibly mixed) modalities (such as text, audio, video, sensor-data etc) and applications ranging from situation monitoring to planning are built on top of the captured 'state' or representation of the real-world. These applications involve querying and analyzing about events and entities that constitute a situation as well as about relationships amongst or between them. In the crisis response domain, for example, various types of sensors at the crisis site, field reports, communication amongst first responders, news reports, eye witness accounts, etc. can be used to extract a representation of the crisis situation, the state of the crisis site, as well as the progress of response activities. Such a representation can then be utilized to build a system to support customized crisis monitoring capabilities. For instance, it could be used to provide a "big picture overview" to the officials at the Emergency Operations Center (EOC) to enable response planning and resource scheduling. Likewise, it could be used to provide localized information of immediate interest to the first-responders on the field in support of search and rescue activities, and to provide up-to-date information to concerned or affected citizens via a community portal. 

Today such situational awareness systems are built in a relatively ad-hoc fashion as applications on top of existing data and knowledge management systems.  Such system, designed as general purpose tools for data (knowledge) management do not support an abstraction suited for representing situations and building situation awareness applications. With the view of overcoming the limitations of existing data management systems, in this paper, we promote an {\em event-centric} approach to modeling and representing situational information.  In very general terms, an event is an occurrence of something of interest of a certain type at a certain place at/over a certain period of time.\footnote{Our view of event is similar in spirit to its definition in the Webster Dictionary where it is defined as a "{\em significant happening}" or an "{\em occurrence}". It is a "{\em fundamental entity of observed physical reality represented by a point designated by three coordinates of place and one of time in the space-time continuum}"}. An event is a semantic, domain-dependent concept, where an event (depending upon the domain) may have associated with it a set of entities that play different roles, and may bear relationships to other events and/or to entities in the real-world. Events, in our view, provide a natural abstraction for modeling, representing, and reasoning about situations.  Not only does the event abstraction provide a natural mechanism/interface for users to query/reason/analyze situational data, it also provides a natural framework to incorporate prior domain or context knowledge in a seamless manner when reasoning about situations (illustrated in more detail later). 

While events are domain and application specific concepts, and their types, properties, and associated entities and relationships may vary from situation to situation, it is also true that despite their differences, events, in general, have certain common properties and relationships (e.g., spatial, and temporal properties), and they support a few common operations and analyses (e.g., spatial and temporal analysis). Exploiting these commonalities, it becomes possible to design a general-purpose {\em event management system} (EMS) that serves as a framework for representing and reasoning about situations. In such an EMS, an event would be treated as a first-class object much in the same way objects and entities are treated in traditional data management systems. 

Our goal in this paper is to explore the viability of a general event management system and to identify challenges in developing such an event management system. We envision such a system (or a system of systems) to support all the components necessary to build event-based situational representations. Specifically, it would support mechanisms to specify domain-specific events, entities, and relationships of interest, provide tools to incorporate domain semantics in reasoning, support languages for querying and analyzing events, as well as mechanisms to indexing and other capabilities to enable efficient data processing. With the above in mind, we enumerate the following requirements for a general purpose Event Management System (EMS).

\begin{enumerate}

\item {\em Situation Modeling Capabilities.} The system should lend itself naturally to modeling events and relationships at an appropriate level of abstraction. Just as schemas or ER diagrams are successful modeling primitives for enterprise structured data, the EMS must provide for appropriate and natural modeling of events and relationships. 

\item
{\em Desired Data Management Capabilities.} An event management system must also have capabilities desired of any data management system, such as simplicity of use, appropriate query language (for events and event relationship oriented  queries in this case), efficient querying and storage, and also interoperability with legacy data sources.

\item
{\em Semantic Representation and Reasoning Capabilities}
An EMS also needs to incorporate {\em domain} and {\em context} knowledge when extracting, representing or reasoning about events. Thus capabilities must be provided to represent such domain and context knowledge (i.e., semantics) as well as utilize it when answering queries about events or relationships between them.

\end{enumerate}

We note that the event management system we seek does not need to be designed from scratch, nor does it need to be designed in isolation from other existing technologies. Many of the required capabilities for EMS have been studied and developed in related areas such as GIS, multimedia and spatio-temporal databases, and data management at large, which can be leveraged. We will discuss the existing literature in this context in Section 2. Designing EMS, however, opens many significant research challenges the foremost of which is identifying an abstraction that captures commonalities across events that can be refined to meet the needs of a large class of situational awareness applications and can also be efficiently realized (potentially using existing data management and knowledge management tools). The key challenge is to develop a generic yet useful model of an event that can be used as a basis of the system design.  Besides the above challenge, there are numerous other technical challenges that arise when representing situational data using events. These include the challenge of {\em disambiguating} events as well as representing their spatial and temporal properties. While prior work exists on disambiguation, and on space and time representation, as will become evident in the paper, many of the solutions/approaches need to be redesigned when representing data at the level of events. 

In the remainder of this paper, we address some of the challenges in developing an EMS. Specifically, we describe a model for events that is guiding our design of an EMS (Section 3), and discuss some of the technical challenges (and solutions we designed) for representing spatial properties of events as well as techniques for event disambiguation (Section 5).

\section{Related Work}

Formal methods for reasoning about events based on explicit representation of events date back at least to work on {\em situation calculus}~\cite{mh69}. Situation calculus treats situations as snapshots of the {\em state} of the world at some time instants. {\em Actions} change one situation to another. These actions are instantaneous, have no duration, and have immediate and permanent effects upon situations. Another formal method is the {\em event calculus}~\cite{ks86} which explicitly represents {\em events} (including actions) that belong to an {\em event type} and generate new situations from old ones. Predicates in event calculus are defined over {\em fluents} which are time-varying property of a domain, expressed by a proposition that evaluates to true or false, depending on the time and the occurrence of relevant events. Predicates on fluents include: Occurs(event,time), HoldsAt(fluent,time), Initiates (event,fluent,time), Terminates (event,fluent,time).

%Hence, the situation calculus is a {\em change-based} approach where the 
%- although events are explicitly represented, they can only be represented as changes in states to a given object.

Reactive systems (including active databases and large system monitoring applications) also explicitly store and reason about events. Here, an event is defined as a system generated message about an activity of interest and it belongs to an event type (situation). In addition to representing events, these applications are also interested in detecting the occurrence of events (e.g.~\cite{spit00,ae04,me04}). Besides detecting and storing {\em primitive} events, these applications detect and represent {\em composite} events which are defined as some sequence of primitive events using an {\em event algebra} (e.g.~\cite{zu99}). As in situation calculus, an event is considered to occur at a precisely determined point in time and has no duration. Although composite events span a time interval they are typically associated with the time-point of the last component event. However, events can be mapped to time intervals to apply queries over the duration of the sequence that a ``compound composite" event matches. All event types consist of core attributes like time point of occurrence, event identifier, event type label, event source identity, and so on. There are a number of commercial products that support such applications including IBM's Tivoli Enterprise Console (and its Common Base Event Infrastructure) and iSpheres EPL Server.
%(http://www.ispheres.com/solutions/ ). 

Our focus is different from this body of work in several ways, namely: 
%(1) This work assumes the environment in which the events occur is relatively static while events we consider occur in a dynamic environment,
(1) Events in reactive systems are well-defined structured messages with restricted variations. In contrast, real-world events are communicated in diverse formats like text, video, audio, etc., (2) Since we deal with real world events, we consider spatial aspects of events which are not dealt with in reactive systems, (3) Information about real-world events can be much more imprecise (as it is derived from potentially noisy source like human reports) and much more complex. (4) Relationships between events (e.g. causation) in reactive systems are typically strong and easier to detect due to the static nature of the environment (system configuration) in which the events occur. Real-world events have weaker relationships and include temporal, spatial and domain relationships. 

Recent work in video content representation has also considered events as foundations of an ontology-driven representation~\cite{nhb04,nzh03,brem04}. The goal of this body of work thus far has been on producing a video event mark-up language that can facilitate data exchange and event recognition. As in situation calculus, an event is defined as a change in the state of an object. A state is a spatio-temporal property valid at a given instant or stable within a time interval. Events can be primitive or composite. {\em Primitive} events are state changes directly inferred from the observables in the video data. Primitive events are more abstract than states but they represent the finest granularity of events. As in situation calculus~\cite{mh69}, time is the critical distinguishing factor between states and events. For example, two identical states with different time values represent two different events. A {\em composite} event is a combination of states and events. Specifically, a composite event is defined by sequencing primitive events in a certain manner. This sequencing can be single-threaded (single-agent based) or multi-threaded (multi-agent based). Events, states and entities can be related to each other using predicates. Spatial and temporal relationships are defined as predicates on members of the time and space domains linked to events. In general this body of work is object-centric, i.e. assumes knowledge of objects precedes knowledge of the event as it defines events as changes in object states. As discussed in Section~\ref{sec:repr-model}, we adopt an event-centric approach. Besides, the constructs in this body of work are tailored to automatic recognition of events from video while we focus on facilitating queries on event data.

%- Existing event modeling approaches (spatio-temporal database techniques) and their associated database technologies (storage and query methods) are object-centric. I.e. events are represented as changes in states of an object. Temporal aspects of events is captured as a set of snapshots of state changes (e.g. work on temporal databases).

%- E.g. Video event representation model (limited in power as it tailored representation of events to automatic recognition of events), Situation awareness ontology, etc.

Event-oriented approaches have also been studied in spatio-temporal data management. The goal here is to represent events associated with geographical/spatial objects. As noted in~\cite{worb05}, the effort on spatio-temporal event representation has evolved in three stages: (1) Temporal snapshot of spatial configurations of events, (2) Object change (captured in terms of change primitives such as creation, destruction, appearance, disappearance, transmission, fission, and fusion) stored as a sequence of past states, and recently (3) Full-fledged representation of changes in terms of events, actions (initiated occurrences), and processes. An example of stage 1 representation is [11] where, starting with an initial state (base map), events are recorded in a chain-like fashion in increasing temporal order, with each event associated with a list of all changes that occurred since the last update of the event vector. The Event-based Spatio-Temporal Data Model (ESTDM)~\cite{pd95} is an example of stage 2. ESTDM groups time-stamped layers to show observations of a single event in a temporal sequence. The ESTDM stores changes in relation to previous state rather than a snapshot of an instance. An event component shows changes to a predefined location (a raster cell) at a particular point in time. The SPAN ontology~\cite{gs05} that defines an event/action/process view and the process calculus based approach of~\cite{worb05} that can also represent event-event relationships are examples of stage 3. Basically, in stage 3, rather than the sequence of past states of each object, the events that caused the state changes are modeled resulting in a a more richer representation. As such, stage 3 can: (1) tell us ``why'' a state exists, and (2) enable us to represent which event caused a state change when multiple events (or sets of events) can potentially cause the same state change.

%- Existing spatio-temporal databases and event models represent the "what" and not the "how". 

\section{Towards an EMS System} \label{sec:repr-model}

This section discusses the issues involved in building an EMS system. Our vision of an EMS is a system that manages events just as a DMBS system manages structured enterprise information.  Thus for an EMS system we are concerned with the (DBMS like) issues of representing event information (modeling), querying and analyzing events, and finally ingesting event information from information sources about situations and events. 
Unlike the concept of records stored in traditional DBMSs, events are a semantically richer concept that lead to some unique challenges. 
For instance, unlike enterprise systems, where the information to be managed is structured and is available in that form as is, event information is embedded in {\em reports} (for instance a text (news) report about a situation or a video (news) coverage) describing or covering situations related to those events. Events need to be {\em extracted} from such reports. Given the extracted event information we may be left with uncertainty about what the information is referring to; for instance (as elaborated on later) there may be ambiguity about some entity referred to in an event and also ambiguity about locations mentioned or referred to. Such uncertainty and ambiguity must be adequately resolved or represented. Furthermore, given that events are a semantic notion, interpreting events, as well as interpreting queries about events requires mechanisms to incorporate domain knowledge and context with both event extraction and querying/reasoning. Querying/analysis techniques on top of EMSs must be able to deal with uncertainty in event descriptions and the corresponding query languages must support constructs to support spatial and temporal reasoning.

\begin{figure}[!t]
\centering
\includegraphics[width=\linewidth]{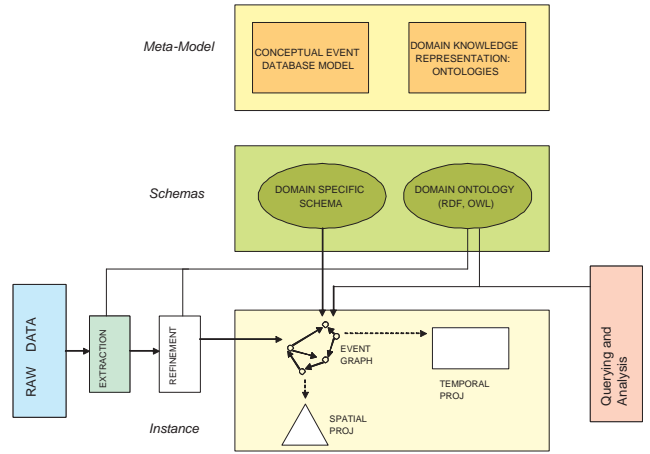}
\caption{Event Management System.}
\label{fig:ems}
\end{figure}

A high-level schematic overview of an EMS system is illustrated in Figure~\ref{fig:ems}. Crucial to the design of an EMS is to develop a model of events. In the following, we discuss some of the key considerations and issues in modeling events, and in developing techniques for querying, analyzing, and extracting and disambiguating events. 

\subsection{Information Modeling}

\begin{figure}[!t]
\centering
\includegraphics[width=\linewidth]{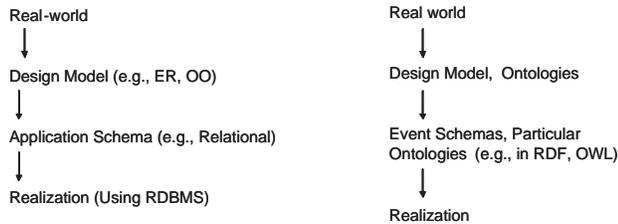}
\caption{DBMS and EMS Systems}
\label{fig:analogy}
\end{figure}

In a DBMS system we start with capturing the real world using design models such as the ER model. We then create application specific schemas, in a particular database model, such as the relational model. Finally there is a physical realization of each database (see Figure ~\ref{fig:analogy}).  In EMS systems too we need to capture the real world (situations and events) in an appropriate design model\footnote{ The work in ~\cite{singh04} proposes some thoughts on an extended ER model for modeling event information}. We also need to pay attention to {\em domain knowledge}, i.e., prior world knowledge that may be related to the events. As we shall explain later, domain knowledge plays a critical role in various facets of an EMS system. Such domain knowledge may be represented in {\em ontologies} which we elaborate on in later sub-sections. We then move to application specific event schemas and application specific domain knowledge (as instantiated ontologies) and finally a physical realization of 
 the information. These levels of abstraction are schematically shown in Figure~\ref{fig:ems}.

We now discuss the various elements and relationships for modeling events. 

\subsection{Building Blocks of the Event Model}

{\bf Report:} A report is the fundamental information source containing event information. A report could be of any modality, for instance an (audio) phone call reporting the event, information in text reports such as text alerts or news stories, or audio-visual information from say a live TV coverage of a situation. A {\em report} is defined then as a physical atomic unit that describes one or more events. 
\\ \\
{\bf Event:} An {\em event} is an instance of an event type in space and time. So an instance of a vehicle having overturned on a road, is an example of an event. A situation comprises of a number of events. Events are extracted from reports. 
\\ \\
{\bf Entity:} An {\em entity} is an object that occupies space and exists for an extended period of time. Events generally have entities, such as people, objects (such as say cars or planes), etc. associated with them. For instance a vehicle overturning event will have the particular vehicle overturned as one of the entities associated with that event. 
\\ \\
{\bf Milieu:} A {\em milieu} is the spatial, temporal or spatio-temporal context in which an event, an object or report is situated. Continuing with the vehicle overturning example, the time and place where the incident occurred are milieus associated with the event.
\\

\begin{figure}[!t]
\centering
\includegraphics[width=\linewidth]{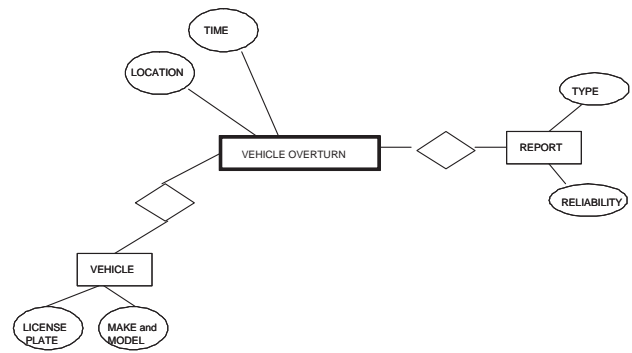}
\caption{Vehicle Overturning Event}
\label{fig:mod1}
\end{figure}

As an example, 
a model for a vehicle overturning event is illustrated in ~\ref{fig:mod1}. The figure illustrates various aspects of the event model introduced so far. It is shown using an ER diagram in the spirit of ~\cite{singh04}. The model captures a "VEHICLE OVERTURN" event, which has associated entities such as "VEHICLE" (the vehicle which overturned) and also one or more "REPORTER" entities which are the person(s) and/or organization(s) reporting that event. 

Notice that space and 
time in the event model should be represented separately from events and objects. This approach is similar to others 
like VERL~\cite{nhb04} and OLAP (where time and space are dimensions). This 
approach (1) allows us to simultaneously use various forms of space and time 
values (i.e. interval/point, region/point, and so on), (2) allows us to 
associate attributes to space (e.g. name, geocode, shape, population) and 
time instances, (3) explicitly represent and reason about relationships 
between space and time instances (for instance using spatial and temporal 
hierarchies). 

%However, notice that instances of space and time are treated 
%differently from event and object instances. Specifically, time and space 
%instances are not assigned unique identifiers that remain the same when 
%their attributes change. Hence, space and time instances are constants.

Events and entities in events may be related amongst or across each other in different ways. Such relationships may be extracted from the reports or inferred from domain knowledge (as described later).
\\

{\bf Relationships}
Each event may be associated with entities as described above, creating the notion of an {\em Event-Entity Relationship.}
The relationships between events and entities can be described as that of 
{\em participation} (an entity participates in an event) or, conversely, 
{\em involvement} (an event involves an entity).
%Case rel -- relationships via an act of participation (between an act and 
an object).
Entities participate in events with a given {\em role}. Examples of kinds of 
these roles include: agentive (object produces, perpetuates, terminates a 
particular event), influencing (facilitate, hindrance), mediating (indirect 
influence)~\cite{gs05,wh04}.
\\

\begin{figure}[!t]
\centering
\includegraphics[width=\linewidth]{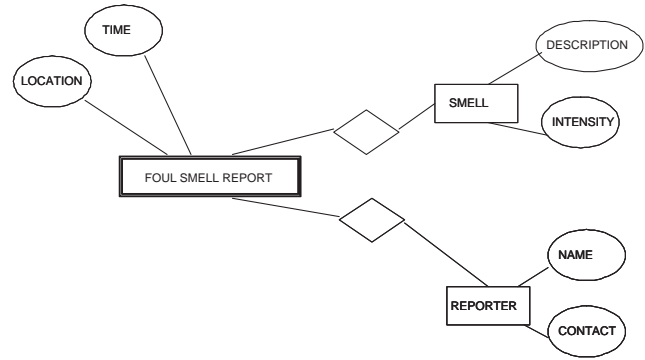}
\caption{Foul Smell Report Event}
\label{fig:mod2}
\end{figure}

Entities may be related to each other. Consider another event, that of a report of a foul smell in the neighborhood with is reported by someone and for which there is a known victim (affected by the smell or the associated gas). A model for this event is illustrated in ~\ref{fig:mod2}. The REPORTER and VICTIM are two entities associated with this event. However the two entities could themselves be related, for instance if the VICTIM is, say, a {\em colleague} of the REPORTER. We then have the notion of an {\em Entity-Entity Relationship.}
\\

Events are related to time and space instances. For instance the vehicle overturning event(referred to as VO) could have occurred at 10:15 am on May 1, 2005 at the intersection of 1rst and Main in Irvine, CA. This gives rise to the notion of an {\em Event-Milieu Relationship}, denoted by types such as: 
$\text{\em at-time}(e,t)$ or
$\text{\em during}(e,T)$,
$\text{\em at-location}(e,L)$,
$\text{\em near-location}(e,L)$, and so on. In these examples, $e$ stands 
for an event, $t$ and $T$ stand for a time point and interval, respectively, 
and $L$ stands for a location. For instance we could instantiate:
$\text{\em at-time}(VO, May 1 2005: 10-15 am)$
\\

%Events can be also related to the clock and 
%calendar~\cite{nhb04}. If the domain of interest contains events that 
%involve changes in spatial relationships over time (e.g. moving points or 
%evolving regions), events are linked to spatio-temporal instances (e.g. 
%represented in a 3D space composed of 2D space and 1D time).
%\\

Locations and times are related amongst themselves as well. Thus we need to capture spatio-temporal or {\em Milieu-Milieu Relationships}.
Relationships between temporal milieu include point-point relationships like 
$\text{\em before}(t_1,t_2)$,
point-interval relationship like $\text{\em begins}(t,T)$, and so on. 
Temporal relationships can also be cyclic (e.g. calendar months) and 
hierarchical (e.g. containment) relationship between intervals or periods. 
Relationships between spatial milieu include part-whole (subsumption) 
relationships, region-region relationships (e.g. touch, disjoint), 
proximity, and so on. The spatial milieu can also be hierarchical where each 
level has its own sets of regions and topological relationships that can be 
used in spatial reasoning. Many temporal relationships or known or can be determined a-priori (for instance a relationship that year 2004 is before year 2005), also spatial relationships, especially between explicitly specified geographic locations can be known a-priori. \\

Events can be related to other events. For instance one event could {\em cause} another event (the vehicle overturning could cause another event, namely that of a road block); an event could {\em hinder} other events (the roadblock could hinder traffic movement in the area). Further, complex events may be {\em composed} of multiple constituent events. Composition of events is important since such composition relationships form hierarchies resulting in composite events. 
\\
Finally, in the situational awareness domain, each event is associated with one or more reports. Thus we have an {\em Event-Report} relationship capturing what report(s) are associated with an event. 

\begin{figure}[!t]
\centering
\includegraphics[width=\linewidth]{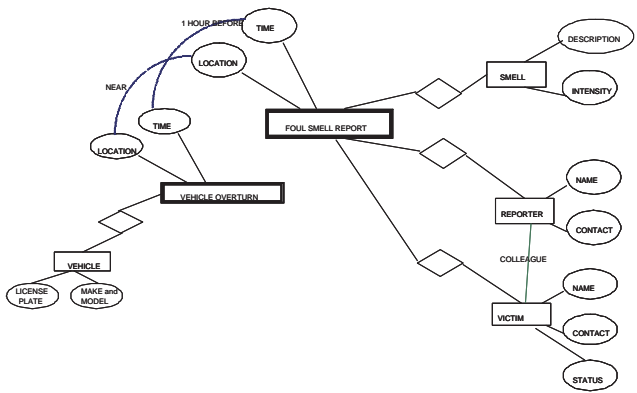}
\caption{Event Relationships}
\label{fig:mod3}
\end{figure}

These relationships are illustrated in ~\ref{fig:mod3}. For instance for the 2 events, the vehicle overturn occurred before the foul smell report so a milieu-milieu relationship (before) between the event times illustrates that. 

\subsubsection{Domain Knowledge}
A data model and instance data can suffice for developing applications, as is typically the case in enterprise information systems. However in situational awareness applications, (prior) {\em domain} or {\em context} knowledge can be valuable as well. For instance in the above example of a foul smell report, it may help to have knowledge about various smells, what chemicals or hazards they may be perpetrated by etc. This holds true in general, for instance for a fire situation it can be valuable to have (precompiled) knowledge about fires and fire-spreads in addition to the information about that particular situation. The same is true for spatial information, as detailed information about places associated with an event is available before the event. 

Thus capturing and representing domain knowledge is an important issue. 
Ontologies are a suitable framework for representing such knowledge. An ontology is defined as a "specification of a conceptualization"~\cite{grub93}. Such ontologies can capture knowledge for a particular domain of interest as well as capture geospatial knowledge of various areas. The Semantic-Web \footnote{ "http://www.w3.org/2001/sw/"} community has devoted 
significant efforts to developing standard, universally accepted, and 
machine processable ontology formalisms in recent years. Common ontology 
representation formalisms, endorsed by the W3C\footnote{ "http://www.w3.org/"} include the Resource 
Description Framework (RDF)~\cite{mill04} and its companion RDF 
Schema(RDFS)~\cite{guha04}, DAML+OIL, OWL~\cite{deb04}, etc. 

\subsection{Querying and Analysis}

Once the event model is determined, an EMS needs to support mechanisms to support retrieval and analysis of events based on their properties and relationships. A natural way to view events is as a network or a graph. We refer to such an event graph as an {\em EventWeb}. The EventWeb is 
an attributed graph where nodes corresponds to events, entities or reports 
 and edges correspond to a variety of 
relationships among events. Both nodes and edges could have associated types that determine the associated attributes. 

Given a graph view, a graph-based query language can be used for querying and analysis for events. We have developed one such graph language named {\em Graph Analytic Language } (GAAL) that extends previous graph languages by supporting aggregation and grouping operators \cite{dawit}. Besides supporting navigational queries (based on path expressions), and selection queries (based on attributes associated with nodes and edges), GAAL also supports the concepts of aggregation and grouping. Using these operators, GAAL can be used to support analytical queries similar in spirit to how OLAP operators are used to support analytical queries over relational data. Such operators allow analytical queries such as queries such as "centrality of a node in a graph", "degree of connectivity between specified nodes" etc. Using GAAL over eventWeb, numerous types of analysis such as causal analysis, dependency analysis, impact analysis etc. can be performed. The aggregation and grouping features of GAAL can also be used to perform such analysis over events at multiple levels of composition/resolution. 

While GAAL as a basis of an event language has a certain appeal for EMS, there are numerous directions in which it will need to be extended to make it suitable for event based systems. First, it needs to be extended to support spatial and temporal reasoning. Space and time (milieu) are integral components of any event based system. Space and time associated with events usually correspond to locations or regions (point /intervals). Since such locations can induce an infinite number of possible spatial and temporal relationships, any one of which could be of interest to the user, such relationships are best not represented as edges in an EventWeb. Instead, an event-based language should support projection of events from and to spatial and temporal dimensions and should seamlessly combine spatial and temporal reasoning along with graph based queries.

\begin{figure}[!t]
\centering
\includegraphics[width=\linewidth]{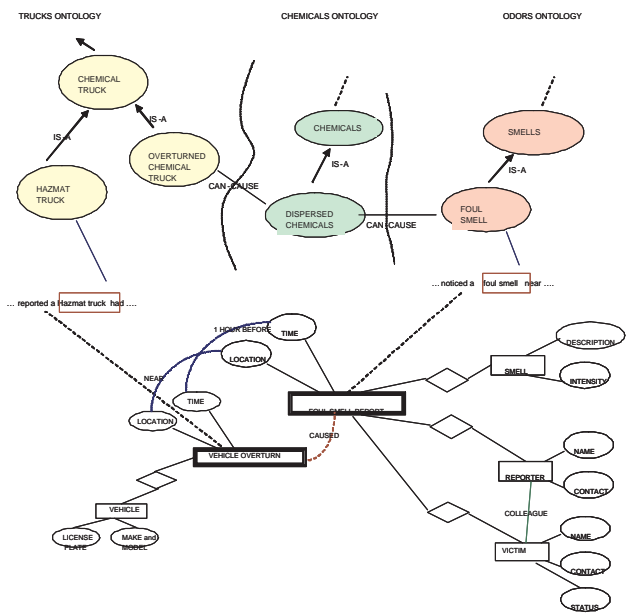}
\caption{Domain Knowledge and Event Relationships}
\label{fig:mod4}
\end{figure}

Another challenge arises due to the semantic nature of events. Unlike languages designed for structured/semi-structured data, where the primary concern is to develop a mechanism to navigate through the structure, semantic associations can play a vital role in expressing and interpreting queries in event based systems.
For instance we may be interested in knowing whether there is any relationship between the vehicle overturn event and the foul smell report event. Just the situational information per-se cannot help us in uncovering such relationships. However incorporation of domain knowledge helps us uncover such possible relationships. For instance the situational information in conjunction with domain knowledge (represented as various ontologies and relationships between the ontologies) enables us to infer that 
{\tt OVERTURNED VEHICLES} which-can-be 
{\tt OVERTURNED-CHEMICAL-TRUCKS} which-can-cause 
\\ {\tt CHEMICAL-DISPERSIONS} which-can-create {\tt FOUL SMELLs}.
This is illustrated in Figure~\ref{fig:mod4}. The language designed for event based system must enable such semantic associations. The work in ~\cite{sheth03} defines the concept of a semantic association over semantically related entities and presents algorithms for extracting such semantic associations between entities. 

Finally, event information is inherently imprecise/uncertain. Events are extracted from reports and extractors might or might not be able to precisely determine the properties of events. Furthermore, there may be ambiguity in the values as well as relationships associated with events (see discussion below).
Such uncertainties must be represented in the basic event model and query language as well as associated query semantics must accommodate for such uncertainties.

\subsection{Data Ingestion}
Unlike enterprise information systems, where the information to be stored is available or entered in required format (i.e., tuples), event information is present in reports of different modalities that talk about or cover a situation containing that event.  Thus events have to be extracted from such reports. 
As mentioned earlier, the extracted event information can have elements of uncertainty and ambiguity. For instance a reference to an entity such as "the vehicle" may be ambiguous and we may need to determine (if possible) which vehicle is the reference to. Finally, uncertainty might be inherent in the description and not resolvable. For instance, spatial location of an event may be specified  as "{\em near} the station". In such a case, query processing techniques to handle such uncertainties must be developed. 

Our progress so far in working towards an event management system has been on techniques of event ingestion. Specifically, we have developed approaches for (1) extraction and representation of spatial properties of events from textual reports and techniques for answering spatial queries using the representation, and (2) domain-independent techniques for disambiguating references and entities associated with events. We are currently developing techniques that exploit domain knowledge (expressed as ontologies) as well as context for information extraction. 

We discuss our work on representing and reasoning about spatial properties of events described in textual reports and also our techniques for disambiguation in the following two section. Such techniques form the building block of an EMS which is our eventual goal.

\section{Handling spatial uncertainty}
\label{sec:uncloc}

As mentioned above, analyzing spatial properties of events is an inevitable part of many decision making and analysis tasks on event data. In our work we have addressed the problem of representing and querying {\em uncertain} location information about real-world events that are described using {\em free text}. As a motivating example, consider (again) the excerpts from two (fictional) transcripts of 911 calls in Orange County (OC): 
\begin{itemize}
\setlength{\itemsep}{0pt}
\setlength{\parskip}{0pt}
 \item {\em \ldots a massive accident involving a hazmat truck on N-I5 between
Sand Canyon and Alton Pkwy. \ldots} 
\item {\em \ldots a strange chemical smell on Rt133 
between I405 and Irvine Blvd. \ldots}
\end{itemize} 
These reports talk about the same event (an accident in this case) that happened at some point-location in Laguna Niguel, CA. However, neither the reports specify the exact location of the accident, nor do they mention Laguna Niguel explicitly. We would like to represent such reports in a manner that enables efficient evaluation of spatial queries and analyses. For instance, the representation must enable us to retrieve accident reports in a given geographical region (e.g., Irvine, Laguna Niguel, which are cities in OC). Likewise, it should enable us to determine similarity between reports based on their spatial properties; e.g., we should be able to determine that the above reports might refer to the same location. 
%it is likely 
%that the two reports above are about the same event given the geography of OC
%(assuming, of course,
%that the time specified in the reports is roughly the same).

Before we describe our technical approach, we briefly discuss our motivation for studying the afore-mentioned problem. We have already alluded to the usefulness of spatial reasoning over free text for 911 dispatch in the example above. We further note that such solutions are useful in a variety of other application scenarios in emergency response. For instance, such a system could support real-time triaging and filtering of relevant communications and reports among first responders (and the public) during a crisis. In our project, we are building Situational Awareness (SA) tools to enable social scientists and disaster researchers to perform spatial analysis over two such datasets: (1) the transcribed communication logs and reports filed by the first responders after the 9/11 disaster, and (2) newspaper articles and blog reports covering the S.E. Asia Tsunami disaster. We believe that techniques such as ours can benefit a very broad class of applications where free text is used to describe events.

Our goal is to represent and store uncertain locations specified in reports in the database so as to allow for efficient execution of analytical queries.
Clearly, merely storing location in the database as free text is not sufficient to answer either spatial queries or to disambiguate reports based on spatial locations. For example, a spatial query such as `retrieve accident reports in the city of Laguna Niguel' will not retrieve either of the reports mentioned earlier. 

To support spatial analysis on free text reports, we need to project the spatial properties of the event described in the report onto the domain $\Omega$. For that, we model uncertain event locations as random variables that have certain probability density functions ({\em pdfs}) associated with them. We develop techniques to map free text onto the corresponding pdf defined over the domain.
% $\Omega$.

Our approach is based on the assumption\footnote{We have validated this claim through a careful study of a variety of crisis related data sets we have collected in the past.} that people report event locations based on certain  {\em landmarks}. Let $\Omega \subset R^2$ be a 2-dimensional physical space in which the events described in the reports are immersed. Landmarks correspond to significant spatial objects such as buildings, streets, intersections, regions, cities, areas, etc. embedded in the space. Spatial location of events specified in those reports can be mapped into {\em spatial expressions} (s-expressions) that are, in turn, composed of a set of {\em spatial descriptors} (s-descriptors) (such as {\tt near}, {\tt behind}, {\tt infrontof}, etc) described relative to landmarks.  Usually, the set of landmarks, the ontology of spatial descriptors, as well as, their interpretation are domain and context dependent. Figure~\ref{table:textNexp} shows excerpts of free text referring to event locations and the corresponding spatial expressions.
\begin{figure}[!hbt]
\small
\centering
\begin{tabular}{l | l}
{\em free text} & {\em s-expression}\\
\hline
`between buildings $A$ and $B$' & $\text{\tt between}(A,B)$\\ 
`near building A'               & $\text{\tt near}(A)$\\
`on interstate I5, near L.A.'   & $\text{\tt within}(\text{I5}) \land \text{\tt near}(L.A.)$\\
%\hline
\end{tabular}
\vspace*{-0.5em}
\caption{Examples of s-expressions.}
\label{table:textNexp}
\end{figure}   
These expressions use $A$ and $B$ as landmarks. While the spatial locations of landmarks are precise, spatial expressions are inherently uncertain: they usually do not provide enough information to identify the exact point-locations of the events.

\begin{figure}[!hbt]
\centering
%\vspace*{-4em}
\includegraphics[width=3in]{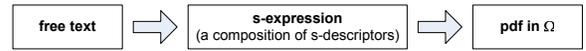}
\vspace*{-0.5em}
\caption{Free text location $\mapsto$ pdf $\in \Omega$.}
\label{fig:text2pdf}
\end{figure}  

Our approach to representing uncertain locations described in free text consists of a two-step process, illustrated in Figure~\ref{fig:text2pdf}. 
First, a location specified as a free-text is mapped into the corresponding s-expression. That in turn is mapped to its corresponding pdf representation. Given such a model, we develop techniques to represent, store and index pdfs so as to support spatial analysis and efficient query execution over the pdf representations.

Our {\bf primary contributions} in this direction are:

\begin{itemize}
\setlength{\itemsep}{0pt}
\setlength{\parskip}{0pt}
\item An approach to mapping uncertain location information from free text into the corresponding
      pdfs in the domain $\Omega$.
\item Methods for  representation and efficient storage of complex pdfs in database.
\item Identification of queries relevant to SA applications.
\item Indexing techniques and algorithms for efficient spatial query processing.
\end{itemize}

\subsection{Modeling location uncertainty}
\label{sec:model}

We model each uncertain location $\ell$, as a continuous random variable (r.v.) which takes values $(x, y) \in \Omega$ and has a certain probability density function (pdf) $f(x, y)$ associated with it. Interpreted this way, for any spatial region $R$, the probability that $\ell$ is inside $R$ is computed as $\int_R f_\ell(x,y) dx dy$. The set of points for which $f(x, y) \not=0$ is called {\em uncertainty region} $U_\ell$ of $\ell$. More specifically, we are interested in conditional density $f(x, y | report)$ which describes possible locations of the event given a particular report. While a report might contain many types of information that can influence $f(x, y | report)$, we concentrate primarily on direct references to event locations, such as `near building A'.
To map locations specified as free text into the corresponding density functions, we employ a divide-and-conquer approach. We first map a free text location into the corresponding s-expression which is a composition of s-descriptors. S-descriptors are less complex than s-expressions, and can be mapped into the corresponding pdfs. The desired pdf for the s-expression is computed by combining the pdfs for s-descriptors. The last step of this process incorporates the prior-distribution into the solution.

%-------------------------------------------
{\bf Mapping free text onto s-expression.}~
Mapping of free-text locations into the s-expressions is achieved by employing spatial ontologies. The development of spatial ontologies is not a focus of our work on spatial uncertainty handling, but we will summarize some of the related concepts in order to explain our approach.

\begin{figure}[!hbt]
\small
\centering
\begin{tabular}{l l l l}
%\hline
{\tt behind}      & {\tt totheleftof}  & {\tt infrontof}  & {\tt near} \\
{\tt between}     & {\tt totherightof} & {\tt withindist} & {\tt within} \\
{\tt indoor}      & {\tt outdoor}      &                  & \\ 
%\hline
\end{tabular}
\vspace*{-0.5em}
\caption{Examples of s-descriptors.}
\label{table:exspdsc}
\end{figure}   

The basic idea is that each application domain $\mathcal{A}$ has, in general, its own spatial ontology $\mathcal{D}(\mathcal{A})$. The ontology defines what constitutes the landmarks in $\mathcal{A}$, and the right way of specifying them in the ontology. It also defines the set of basic s-descriptors $\{\mathcal{D}_1,\mathcal{D}_2, \ldots,\mathcal{D}_n\}$, such that any free-text location from $\mathcal{A}$ can be mapped onto a composition of s-descriptors.
Examples of s-descriptors are provided in Figure~\ref{table:exspdsc}. Each s-descriptor is of the form $\mathcal{D}_i(\mathcal{L}_1,\mathcal{L}_2,\ldots,\mathcal{L}_m)$: it takes as input $m \in N$ landmarks, where $m$ is determined by the type of s-descriptor and can be zero. For instance, Figure~\ref{table:textNexp} shows some free text referring to event locations and the corresponding spatial expressions.
Some s-descriptors may not take any parameters. For instance, an ontology may use the concept of {\tt indoor} and  {\tt outdoor}, to mean `in some building' and `not in any building' respectively.

We have addressed the most common type of s-expression: {\tt AND}-expressions. 
Another type of an s-expression is an {\tt OR}-expression, but it rarely arises in practice. An expression of type {\tt AND} arises when the same location
$\ell$ is described using $n$ different descriptions $s_1,s_2,\ldots,s_n$, which we denote as $s = s_1 \land s_2 \land \cdots \land s_n$. Here $s_1,s_2,\ldots,s_n$ are subexpressions of $s$. As an example of an {\tt AND}-expression, assume a person is asked `where are you?' to which he replies `I am near building $A$ {\em and} near building $B$', which corresponds to the s-expression $\text{\tt near}(A) \land \text{\tt near}(B)$. 
%

\begin{figure}[!hbt]
\centering
\includegraphics[width=\linewidth]{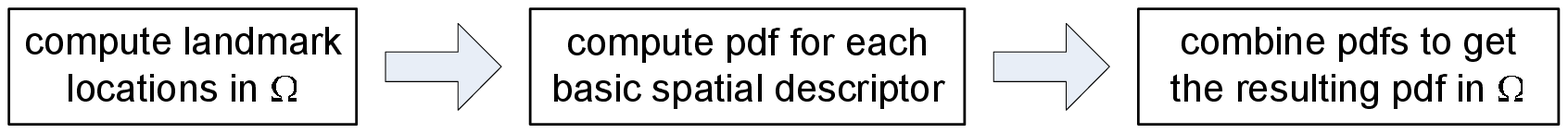}
\vspace*{-2em}
\caption{Combination of s-descriptors $\mapsto$ pdf $\in \Omega$.}
\label{fig:sdsc2pdf}
\end{figure}

%---------------------------------------------
{\bf Pdf for a single s-descriptor.}~
We observe that merely representing locations as spatial expressions is not sufficient. We also need to be able to {\em project} the meaning of
each s-expression onto the domain $\Omega$. We achieve that by (a) being able to compute the projection (i.e., the pdf) of each individual s-descriptor in the s-expression; and (b) being able to combine the projections. This process is illustrated in Figure~\ref{fig:sdsc2pdf}.

We first describe how a basic s-descriptor can be projected into $\Omega$ in an automated fashion. Then, we will demonstrate how to compose those projections to determine the pdfs for s-expressions.
It is important to note that our overall approach is {\em independent from
the algorithms for mapping basic s-descriptors into pdfs}.

To illustrate the steps of the algorithm more vividly, consider a simple 
\begin{figure}[!hbt]
\small
\null
\begin{minipage}[t]{0.49\linewidth}
\centering
\includegraphics[width=\linewidth]{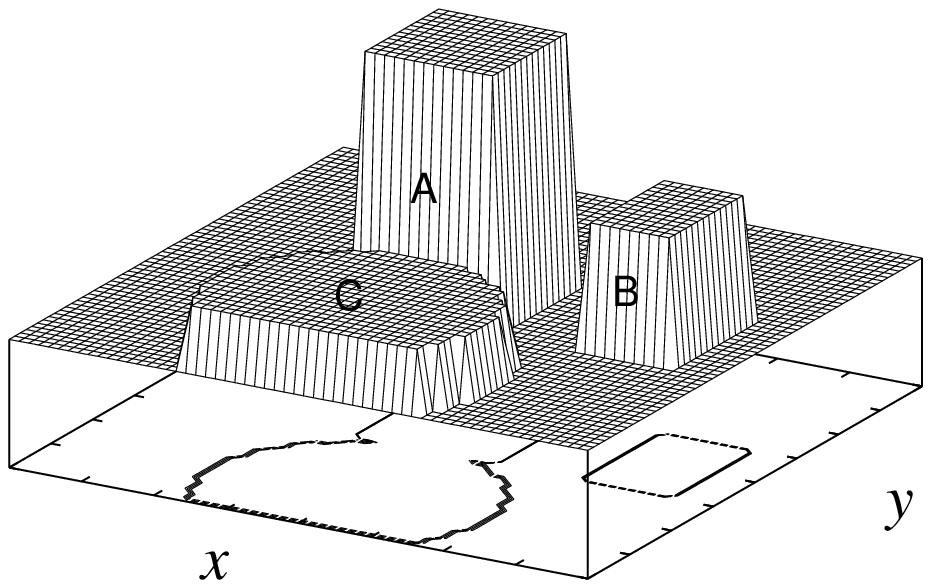}\\
(a) Part of campus.
\end{minipage}
\hfill
\begin{minipage}[t]{0.49\linewidth}
\centering
\includegraphics[width=\linewidth]{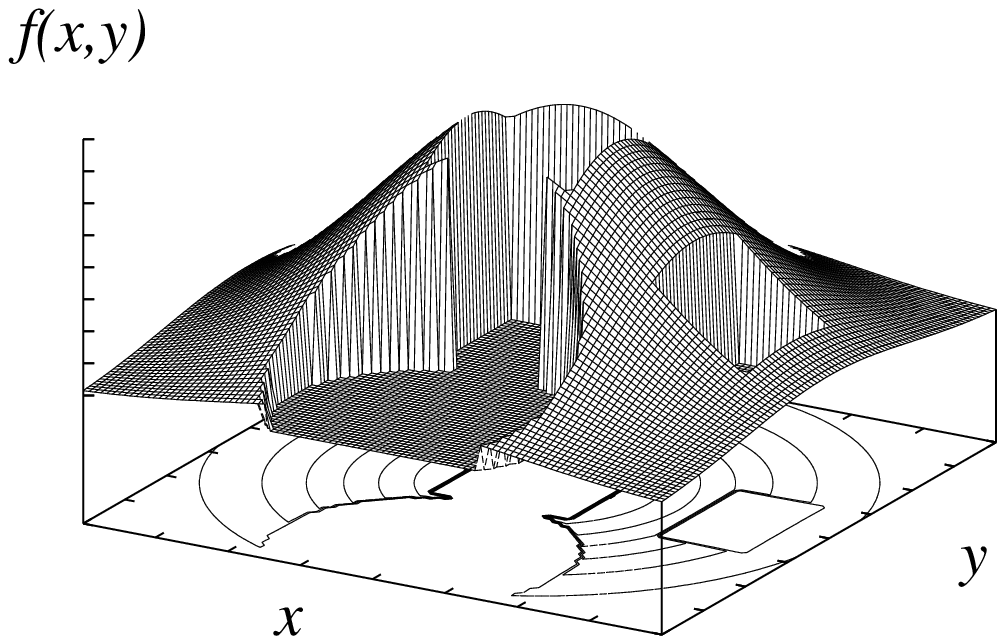}\\
(b) $\text{\tt outdoor} \land \text{\tt near}(A)$.
\end{minipage}
\null
\vspace*{-0.5em}
\caption{Buildings on a campus and various pdfs.}
\label{fig:bldgpdf}
\end{figure}  
scenario demonstrated in Figure~\ref{fig:bldgpdf}(c). This figure shows a portion of a university campus with three buildings $A$, $B$, and $C$. Assume a person reports an event that happened at location $\ell = \text{\tt outdoor} \land \text{\tt near}(A)$. 

The method we use for modeling the pdf \\
$f(x,y | \mathcal{D}(\mathcal{L}_1,\mathcal{L}_2,\ldots,\mathcal{L}_m))$ for any s-descriptor $\mathcal{D}(\mathcal{L}_1,\mathcal{L}_2,\ldots,\mathcal{L}_m)$ requires making reasonable assumptions about the functional form of \\
$f(x,y | \mathcal{D}(\mathcal{L}_1,\mathcal{L}_2,\ldots,\mathcal{L}_m))$. The model depends on the nature of each descriptor, and the spatial properties of the landmarks it takes as input, such as the size landmark footprints, their heights. The model is calibrated by learning the parameters from data. The modeling assumptions can be refined or rejected later on, e.g. using Bayesian framework. 

For instance, for s-descriptor {\tt outdoor} we can define the pdf $f(x,y|{\tt outdoor})$ as having the uniform distribution everywhere inside the domain $\Omega$ except for the footprints of the buildings that belong to $\Omega$. That is $f(x,y) = c$ for any point $(x,y) \in \Omega$ except when $(x,y)$ is inside the footprint of a building, in which case $f(x,y) = 0$. The real-valued constant $c$ is such that $\int_\Omega f(x,y) dx dy = 1$.

Another example is an s-desriptor of $\text{\tt near}(A)$ which means somewhere close to the landmark $A$ (the closer the better), but not inside $A$. Let us note that, unlike the density for {\tt outdoor}, the real density for $\text{\tt near}(A)$ clearly is not uniform. Rather, a more reasonable pdf can be a variation of the truncated-Gaussian density, centered at the center of the landmark, with variance determined by the spatial properties of the landmark $A$ (its height, the size of its footprint). 
Also, since the location cannot be inside $A$, the values of that density should be zero for each point inside the footprint of the landmark. This way we can determine the pdf for each instantiated s-descriptor in an automated, non-manual fashion.

%--------------------------
{\bf The pdf of a spatial expression.}~
We have developed formulas for computing the pdfs for {\tt AND}-expressions, by being able to combine the pdfs of the underlying basic s-descriptors. Figure~\ref{fig:bldgpdf}(b) illustrates an example of a pdf for the s-expression $\text{\tt outdoor} \land \text{\tt near}(A)$, evaluated
in the context of the scenario in Figure~\ref{fig:bldgpdf}(a).
Note that in SA domains {\bf pdfs might have very complex shapes}, significantly more involved than those traditionally used. Thus we devise special methods for representing and storing pdfs.

%===============================================================================
\subsection{Spatial Queries}
\label{sec:qry}

SA applications require support of certain types of queries, the choice of which is motivated by several factors. The three salient factors are: the necessity, triaging capabilities, and quick response time. The necessity factor means determining which core types of spatial queries (e.g., range, NN, etc) are necessary to support common analytical tasks in such applications. In crisis situations {\em triaging capabilities} can play a decisive role by reducing
the amount of information the analyst should process. Those capabilities operate by restricting the size of query result sets and filtering out, or, {\em triaging}, only most important results, possibly in a ranked order. Similarly, the solutions that achieve quick query response time, perhaps by sacrificing other (less important) qualities of the system, are required to be able to cope with large amounts of data.

\begin{figure}[!hbt]
\centering
%\vspace*{-1em}
\includegraphics[width=0.9\linewidth]{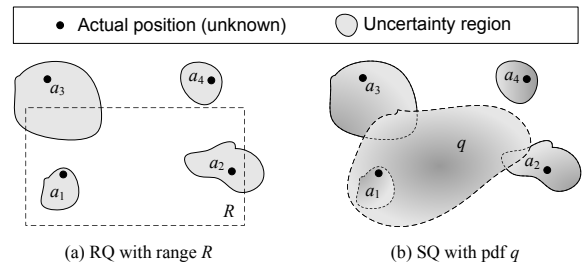}
%\vspace*{-2em}
\vspace*{-1em}
\caption{Examples of RQ$(R)$ and SQ$(q)$.}
\label{fig:RQSQ}
\end{figure}  

We have studied extensively two fundamental types of queries, which must be supported by SA applications: region and similarity queries, illustrated in Figure~\ref{fig:RQSQ}. We have designed and evaluated several modifications of those basic types of queries, which support triaging capabilities and allow for more efficient query processing. Specifically, we have developed algorithms for efficient evaluation of the threshold and top-$k$ versions of those queries.

\begin{comment}
Consider the example in Figure~\ref{fig:RQSQ}(a). If we assume the locations of elements $A=\{a_1,a_2,a_3,a_4\}$ are uniformly distributed in the specified (shaded) uncertainty regions, then the probabilities of these elements of being inside $R$ might be $1.0, 0.7, 0.4, 0.0$ respectively. Then the result set for the corresponding RQ is $\{a_1,a_2,a_3\}$.
\end{comment}

\subsection{Representing and indexing pdfs}
\label{sec:repr}

%---------------------------------------------------------
{\bf Histogram representation of pdfs.}~
In order to represent and manipulate pdfs with complex shapes, we first quantize the space by viewing the domain $\Omega$ as a fine uniform grid $G$ with cells of size $\delta \times \delta$. The grid $G$ is {\em virtual} and is never materialized. Using this grid we then convert the pdfs into the corresponding histograms. That is, for the pdf $f_\ell(x,y)$ of a location $\ell$ we compute the probability $p_{ij}^\ell$ of $\ell$ to be inside cell $G_{ij}$, i.e. $p_{ij}^\ell=\int_{G_{ij}} f_\ell(x,y) dx dy$. The set of all $p_{ij}^\ell \not= 0$ defines a histogram for $\ell$.

\begin{figure}[!hbt]
\null
\begin{minipage}[t]{0.49\linewidth}
\centering
\includegraphics[width=\linewidth]{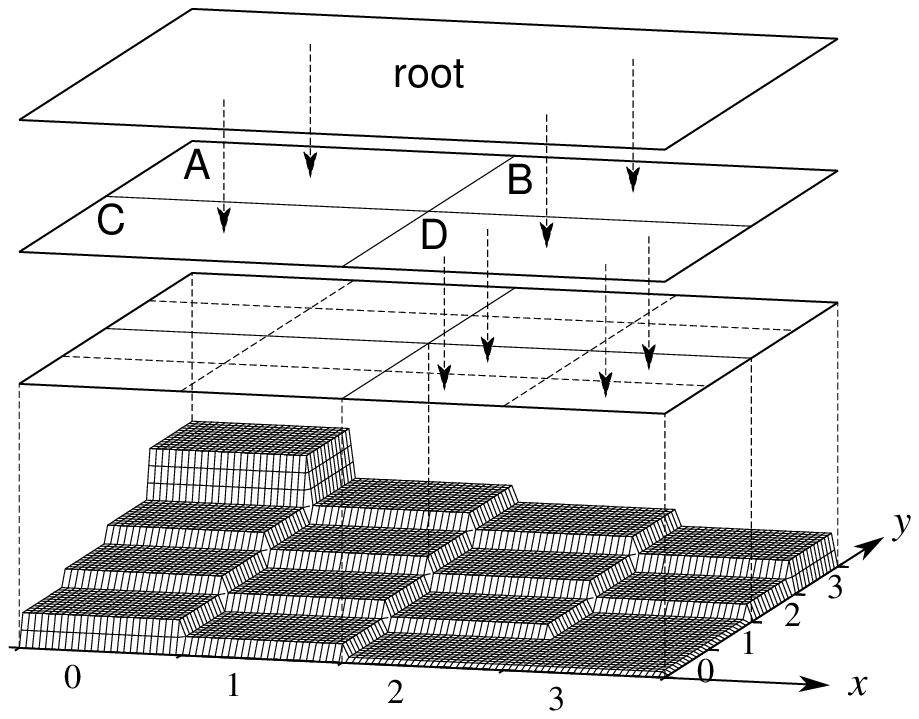}\\
(a) Before compression.
\end{minipage}
\hfill
\begin{minipage}[t]{0.49\linewidth}
\centering
\includegraphics[width=\linewidth]{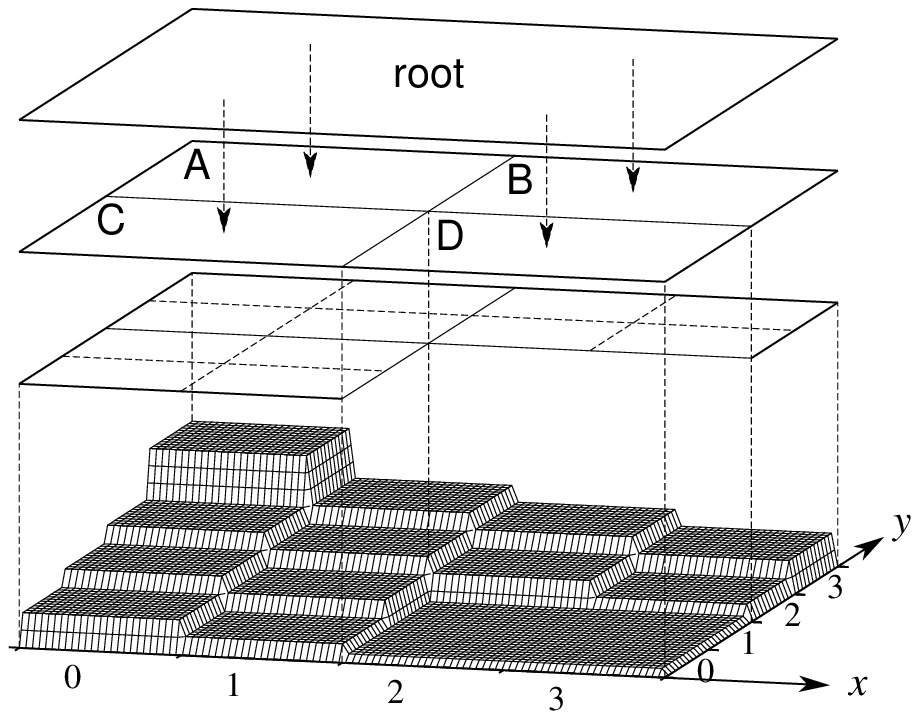}\\
(b) After compression.
\end{minipage}
\null
\vspace*{-0.5em}
\caption{Quad-tree representation of pdf.}
\label{fig:qtree}
\end{figure}  

When we approximate a pdf with a histogram representation, we loose the information about the precise shape of the pdf, but in return we gain several advantages. The main advantage is that manipulations with pmfs are less computationally expensive than the corresponding operations with pdfs, which involve costly integrations. The latter is essential, given that SA applications require quick query response time, especially in crisis scenarios. Therefore, the loss due to approximation and the advantages of using pmf should be balanced to achieve a reasonable trade-off.

%---------------------------------------------------------
{\bf Quad-tree representation of pdfs.}~
We further improve the histogram representations of pdfs by indexing histograms with quad-trees. First we build a {\em complete} quad-tree $\mathcal{T}_\ell$
for each histogram $H_\ell$. Each node $\mathcal{N}$ in the resulting quad-tree $\mathcal{T}_\ell$ stores certain aggregate information that allows for efficient query processing. We have explored several quad-tree (lossy) compression algorithms that trade accuracy of representation for efficiency of query processing.

%---------------------------------------------------------
{\bf Indexing quad-trees with a grid.}~
Assume the goal is to evaluate a $\tau$-RQ with some threshold. The quad-tree representation of pdfs might help to evaluate this query over each {\em individual} location $\ell \sim f_\ell(x,y)$ stored in the database faster. However, if nothing else is done, answering this query will first require a {\em sequential scan} over all the locations stored in the database, which is undesirable both in terms of disk I/O as well as CPU cost. 
To solve this problem we can create a {\em directory index} on top of $U_\ell$ (or, MBR of the histogram) for each location $\ell$ in the database. We have designed a specific grid-based index for this goal and demonstrated its superiority over traditional techniques by extensive empirical evaluation.

We have extensively studied the proposed approach empirically. In our experiments, we use a real geographic dataset, which covers $4 \times 4$ km$^2$ the New York, Manhattan area. The uncertain location data was derived based on the 164 reports filed by Police Officers who participated in the events of September 11, 2001. The number of the reports is rather small for testing database solutions; hence we have generated {\em synthetic} datasets of s-expressions, based on our analysis of the reports. The experimental study showed the feasibility and the efficiency of the proposed approach as well as its superiority over existing techniques.

\section{Event disambiguation}
\label{sec:disamb}

The area of information quality studies various problems that arise when raw datasets must be converted to a normalized representation so that they can be analyzed by various applications. The same problems are unavoidable when the information from raw reports, especially from textual information created by humans, must be processed to create event representations.. The problem with data can arise at all levels of event representation: 
(1) at the attribute level, the values of the attributes of events/relationships can be incomplete, uncertain, erroneous, or missing;
(2) at the event level, duplicate events might exist in the database;
(3) at the relationship level, due to uncertain description of events, there might be uncertainty in how a relationship/edge should be created in the EventWeb, i.e. which entities this edge should connect.

Event disambiguation is the task of creating accurate event representations from raw datasets, possibly collected from multi-modal data sources. In the following discussion we will focus primarily on the event disambiguation challenges that are related to deduplication. Those challenges arise mainly because the information might be compiled from different data sources, which may describe the same events. A good example is news reports, which often describe the same event. Another example is people calling in a 911 center to report an accident: a major accident is typically reported multiple times by different people, putting a strain on 911 centers. Detecting duplicates is important in this context for proper resource planning and response. 

In fact, removing duplicates in datasets is one of the key driving forces behind the whole research area of information quality. The reason is that (a) the problem is common in datasets; and (b) duplicate data items often negatively affect data mining algorithms, which produce wrong results on non-deduplicated data. The approaches for solving such problems with data are classified into
{\em domain-specific} or {\em domain-independent} categories. 
Since we are interested in applying our algorithm to a variety of SA domains, we will be looking into domain-independent techniques.

The disambiguation problem is challenging because the same event can be described very differently in different data sources, and even in a single data source. Traditional domain-independent cleaning techniques rely on analyzing event {\em features} for disambiguation, hence we refer to them as the {\em feature-based similarity (FBS)} methods. They measure the degree of similarity of two events by first computing the similarity of their attributes and then combining those attribute similarities into overall similarity of the two events.
However, there is additional information often available in datasets, which is not explored by traditional techniques. This information is in the {\em relationships} that exist among entities stored in the dataset. An analysis of the {\em connection strength} $c(u,v)$ between two entities $u$ and $v$, stored in the relationship chains between them, can help to decide whether $u$ and $v$ refer to the same entity or not.

%--------------------- 
\subsection{Disambiguation problems}

\begin{figure}[!hbt]
\centering
\includegraphics[width=0.9\linewidth]{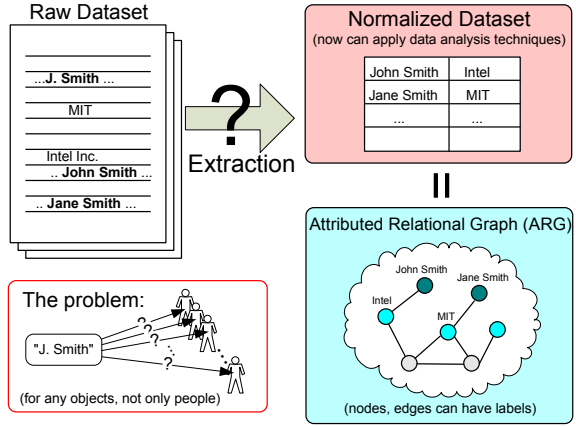}
\caption{Event disambiguation.}
\label{fig:RelDC1}
\end{figure}  

The generic framework we are currently developing, called {\em Relationship-based Event Disambiguation (RED)}, can help solve various disambiguation challenges. One of those challenges is illustrated in Figure~\ref{fig:RelDC1},
where the problem is formulated as follows. When processing an incoming event, the application may determine that the event is already stored in the database. However, the description of the event might be such that it matches the descriptions of multiple stored events, instead of a single one. The goal is, for the event being processed, to identify the right matching event, stored in the database. This problem is fairly generic, and it arises not only for event data. In Figure~\ref{fig:RelDC1} this point is illustrated by showing that the goal is, for a description ``J. Smith", to determine to which particular ``J. Smith" it refers to, in the given context.

Another challenge that RED can help to solve is to deduplicate the same events from the dataset. That is, given that events in the dataset are represented via descriptions, the goal is to consolidate all the descriptions into groups. The ideal consolidation is such that each resulting group is composed of the event descriptions of just one event, and all the descriptions of one event are assigned to just one group.

%--------------------- 
\subsection{RED approach}

\begin{figure}[!hbt]
\centering
\includegraphics[width=0.9\linewidth]{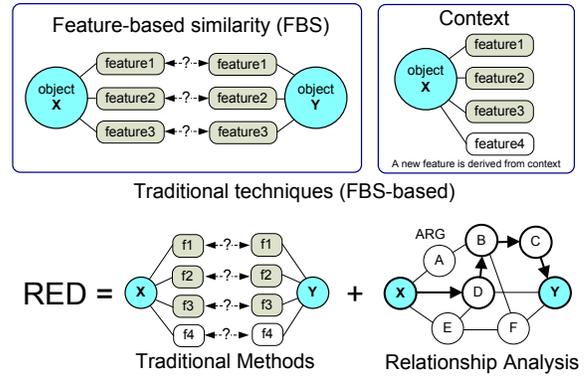}
\caption{Traditional methods vs. RED approach}
\label{fig:RelDC2}
\end{figure}  

Figure~\ref{fig:RelDC2} illustrates the difference between the traditional feature-based domain-independent data cleaning techniques and RED. The traditional techniques, at the core, rely on analyzing object features. Those features are either standard/regular features of the objects, or the ``context attributes" -- the features derived from the context, which a few the recent techniques might be able to employ. RED however proposes to enhance the core of those approaches, by considering relationships that exists among entities.

To analyze relationships, the approach views the underlying dataset as an attributed relational graph (ARG). The nodes in this graph represent entities and the edges represent relationships. The analysis is based on what we refer to as the Context Attraction Principle (CAP). The CAP is a hypothesis, which has been proven empirically for various datasets. Simply put, it states that if two entities $u$ and $v$ refer to the same object, then the connection strength between their context is strong, compared to the case where $u$ and $v$ refer to different objects.

%------------------------
\subsubsection{Reference disambiguation}

To solve the first disambiguation challenge identified above, known as
{\em reference disambiguation}, RED introduces the new concept of a {\em choice} node, illustrated
\begin{figure}[!hbt]
\centering
\includegraphics[width=1in]{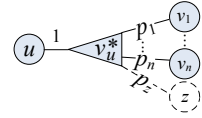}
\vspace*{-1em}
\caption{Choice node.}
\label{fig:choice}
\end{figure}  
in Figure~\ref{fig:choice}. A choice node is created to represent an uncertain relationship. In Figure~\ref{fig:choice} the choice node \choice{u} is created for a relationship between objects $u$ and $v$. However, the description of object $v$ is such that it matches objects $v_1,v_2,\ldots,v_n$ and there is also a possibility that the described object is not in the database, denoted by a virtual object $z$. The choice node \choice{u} represents the fact that $u$ is connected to either one $v_1,v_2,\ldots,v_n$ or $z$. That is if there were no uncertainty and we knew the ground truth, only one edge would exist i.e.,  between $u$ and $v$. The goal is to decide which $v_i$ is $v$. The edges between \choice{u} and $v_1,v_2,\ldots,v_n,z$ have weights associated with them. The weight $p_i$ is a real number between zero and one, which reflects the degree of confidence that $v$ is $v_i$. All weights sum to 1. The goal is then to assign those weights. As the final step, the algorithm picks $v_i$ with the largest $p_i$ as $v$.

To assign those weights the algorithm discovers relationships that exist between $u$ and each $v_i$. It then uses a $c(u,v)$ model to compute the connection strength $c(u,v_i)$ stored in the discovered relationships, for each $v_i$.
Since in general the discovered relationships can go via choice nodes, $c(u,v_i)$ returns an equation that relates $c(u,v_i)$ to the edge weights of other choice nodes (rather than a scalar value). The algorithm then constructs another equation that relates $p_i$ and $c(u,v_k)$ for all $v_k$. 
This procedure is repeated for each choice node in the ARG. In the end the algorithm maps the disambiguation problem into the corresponding optimization problem, which can be solved either using an of the shelf math solver or iteratively.

Note that the algorithm does not process entities sequentially, but rather solves the problem for all the objects simultaneously, by resolving the corresponding optimization problem. 

%------------
\subsubsection{Connection strength models}
Recently there has been a spike of interest by various research communities in the measures directly related to the $c(u,v)$ measure. Since the $c(u,v)$ measure is at the core of the proposed RED approach, we next analyze several principal models for computing $c(u,v)$. For brevity, we limit the discussion to the models that have been employed in our work. For some of these models, we use only their semantic aspects, while procedurally we compute $c(u,v)$ differently.

{\bf Diffusion Kernels.}~
The earliest work in this direction that we can trace is in the area of kernel-based pattern analysis \cite{kernel_book}, specifically the work on the diffusion kernels, which are defined below.

A {\em base similarity graph} $G=(S,E)$ for a dataset $S$ is considered. The
vertices in the graph are the data items in $S$. The undirected edges in this graph are labeled with a `base' similarity $\tau({\bf x}, {\bf y})$ measure. That measure is also denoted as $\tau_1({\bf x}, {\bf y})$, because only the direct links (of size 1) between nodes are utilized to derive this similarity. The base similarity matrix ${\bf B}={\bf B}_1$ is then defined as the matrix whose elements ${\bf B}_{\bf xy}$, indexed by data items, are computed as ${\bf B}_{\bf xy} = \tau({\bf x}, {\bf y}) = \tau_1({\bf x}, {\bf y})$. Next, the concept of base similarity is naturally extended to paths of arbitrary length $k$. To define $\tau_k({\bf x}, {\bf y})$, the set of all paths $P_{\bf xy}^k$ of length $k$ between the data items ${\bf x}$ and ${\bf y}$ is considered. The similarity is defined as the sum over all these paths of the products of the base similarities of their edges: 
\[
\tau_k({\bf x}, {\bf y}) = 
\sum_{({\bf x}_1{\bf x}_2\ldots{\bf x}_k)\in P_{\bf xy}^k}
\prod_{i=1}^k \tau_1({\bf x}_{i-1}, {\bf x}_i)
\]
Given such $\tau_k({\bf x}, {\bf y})$ measure, the corresponding similarity matrix ${\bf B}_k$ is defined. It can be shown that ${\bf B}_k = {\bf B}^k$.
The idea behind this process is to enhance the base similarity by those indirect similarities. For example, the base similarity ${\bf B}_1$ can be enhanced with similarity ${\bf B}_2$, e.g by considering a combination of the two matrices: ${\bf B}_1 + {\bf B}_2$. The idea generalizes to more then two matrices. For instance, by observing that in practice the relevance of longer paths should decay, it was proposed to introduce a decay factor $\lambda$ and define what is known as the {\em exponential diffusion kernel}:
$
{\bf K} = \sum_{k=0}^\infty \frac{1}{k!} \lambda^k {\bf B}^k = \exp(\lambda {\bf B}).
$
The {\em von Neumann diffusion kernel} is defined similarly:
$
{\bf K} = \sum_{k=0}^\infty \lambda^k {\bf B}^k = ({\bf I} - \lambda{\bf B})^{-1}.
$
The diffusion kernels can be computed efficiently by performing eigen-decomposition of ${\bf B}$, that is ${\bf B} = {\bf V}' {\bf \Lambda} {\bf V}$, where the diagonal matrix ${\bf \Lambda}$ contains the eigenvalues of {\bf B}, and by making an observation that for any polynomial $p(x)$, the following holds
$p({\bf V}' {\bf \Lambda} {\bf V}) = {\bf V}' p({\bf \Lambda}) {\bf V}$. The elements of the matrix ${\bf K}$ exactly define what we refer to as the connection strength: $c({\bf x}, {\bf y}) = {\bf K}_{\bf xy}$.

The solutions proposed for the diffusion kernels work well, if the goal is to 
compute $c(u,v)$ for all the elements in the dataset. They are also very useful for illustration purposes. However in data cleaning the task is frequently to compute only some of $c(u,v)$'s, thus more efficient solutions are possible. 
Also, often after computing one $c(u,v)$, the graph is adjusted in some way, which affects the values of the rest of $c(u,v)$'s, computed after that.

{\bf Random walks in graphs.}~
Another common model used for computing $c(u,v)$ is to compute it as the probability to reach node $v$ from node $u$ via random walks in the graph. That model has been studied extensively, including in our work on reference disambiguation \cite{SDM05::dvk,TR}.

{\bf Parameterizable models.}~
In the context of data cleaning the existing techniques have several disadvantages. One disadvantage is that the true `base' similarity is rarely known in real-world datasets. Some existing techniques try to mitigate that by {\em imposing} a similarity model. However, the CAP principle implies {\em its own} similarity measure, and any imposed model, created for its own sake in isolation from the specific application, might have little to do with it. Ideally, the similarity measure should be derived directly from data for the specific application at hand that employs it. One step toward achieving this, is to consider {\em parameterizable} models and then try to learn an optimal combination of parameters directly from data. We have explored such an approach in \cite{PODS05::dvk2} for the problem of reference disambiguation. The model is somewhat similar to that of the diffusion kernels but where certain base similarities $\tau({\bf x}, {\bf y})$ are initially specified a as weight-variables, which are learned later directly from data.

%-----------------------------------------------
\subsubsection{Object consolidation}

The second challenge that RED solves is known as {\em object consolidation}.
The goal of object consolidation is to accurately group the representations of the same objects together. Our ongoing work solves this challenge by combining feature based similarity with analysis of relationships, in a similar manner to the solution proposed for reference disambiguation. However, the overall problem is different from that of reference disambiguation and solved by employing graph partitioning algorithms \cite{IQIS05::dvk}.

\section{Conclusion}
\label{sec:conc} 

In this paper we proposed our vision of an Event Management System (EMS) as a platform for building Situational Awareness (SA) applications, with events as the fundamental abstraction that comprise situations. We discussed several aspects of an EMS, such as information modeling, querying and analysis, and data ingestion, and also presented our work on location uncertainty reasoning and event disambiguation in more detail.

\bibliographystyle{abbrv}
\bibliography{bib/dc,bib/dvk,bib/event,bib/eventdb,bib/semanticsdb,bib/linkmining,bib/loc,bib/related-stella,bib/solver.bib} 

\end{document}